\documentclass[mathleft
]{an}
\usepackage{graphicx}
\usepackage{times}
\begin{document}

% The following seven commands are intended for editorial usage and should be ignored by
% the author(s).
\Pagespan{1}{}% Document's page range. 
% If second parameter is left empty, the last page is computed automatically.
\Yearpublication{2012}%
\Yearsubmission{2012}%
%\Month{0}%   
%\Volume{00333}%  
%\Issue{5/6}% 
% \DOI{10.1002/asna.201211698}% 
\DOI{}

\title{The Galactic thin and thick disk}

\author{M. Steinmetz\inst{}\thanks{Corresponding author: {msteinmetz@aip.de}}
}
\titlerunning{The Galactic thin and thick disk}
\authorrunning{M. Steinmetz}
\institute{
Leibniz-Institut f\"ur Astrophysik Potsdam (AIP), An der Sternwarte 16, D-14482 Potsdam, Germany
}

\received{2012 Apr 30}
\accepted{2012 Apr 30}
%\publonline{2012 Jun 15}


\abstract{The ongoing large spectroscopic surveys of the Milky Way such as SEGUE and RAVE have enabled us to take a fresh look at the structure of the Galactic thin and thick disks, and how their structure fits within the framework of structure formation via hierarchical clustering. In this article I will summarize some recent results mainly based in the RAVE survey with respect to the structure of the Galactic disks, their possible origin as well as indications of substructure and asymmetries.}

\maketitle

\section{Introduction}
Over the past decades it has been commonly assumed that galactic disks are fairly fragile objects that are easily perturbed and puffed up by encounters with other galaxies, even those of low mass (Lacey \& Ostriker 1985).
Considering the large abundance of galaxy disks as the dominant galaxy population at luminosity similar to that of the Milky Way, the pure presence of disks seem to provide strong constraints on the accretion history of the Milky Way and of other galaxies (Toth \& Ostriker 1992). Furthermore, in a recent study by Kormendy et al.~(2010), about 11 out of 19 nearby spirals with ${V_{\rm circ}>150}$ km\,s$^{-1}$ are consistent with having no classical bulge or bar component whatsoever - both features are seen as indicators of encounters with larger satellite galaxies during the formation history of a galaxy. Consequently, a common conclusion sees the formation history of the Galactic disks as fairly quiescent.

These findings seem to be in strong contrast to the cosmological picture of
galaxy formation, in which galaxies like the Milky Way form in a series of
merging events. Indeed, forming bulge-less disk galaxies has been notoriously
difficult in cosmological simulations (Piontek \& Steinmetz 2011; Scannapieco et
al. 2011), even though there has been considerable progress in the past few
years (Guedes et al. 2011; Brook et al. 2011). As far as the halo of our Milky
Way is concerned, the recent years have witnessed a dramatic increase in the
number of identified substructure in the stellar halo (e.g. Belokorov et al. 2006), consistent with the conclusion that the stellar halo of the Milky Way is built exclusively out of debris from disrupted satellites (Bell et al. 2008). Substructure seems also to be ubiquitous in the outskirts of M31 (Richardson et al. 2008). The data of large imaging surveys, e.g., the SDSS or the ACS on HST have been critical for these findings.

\begin{figure}
\includegraphics[width=80mm]{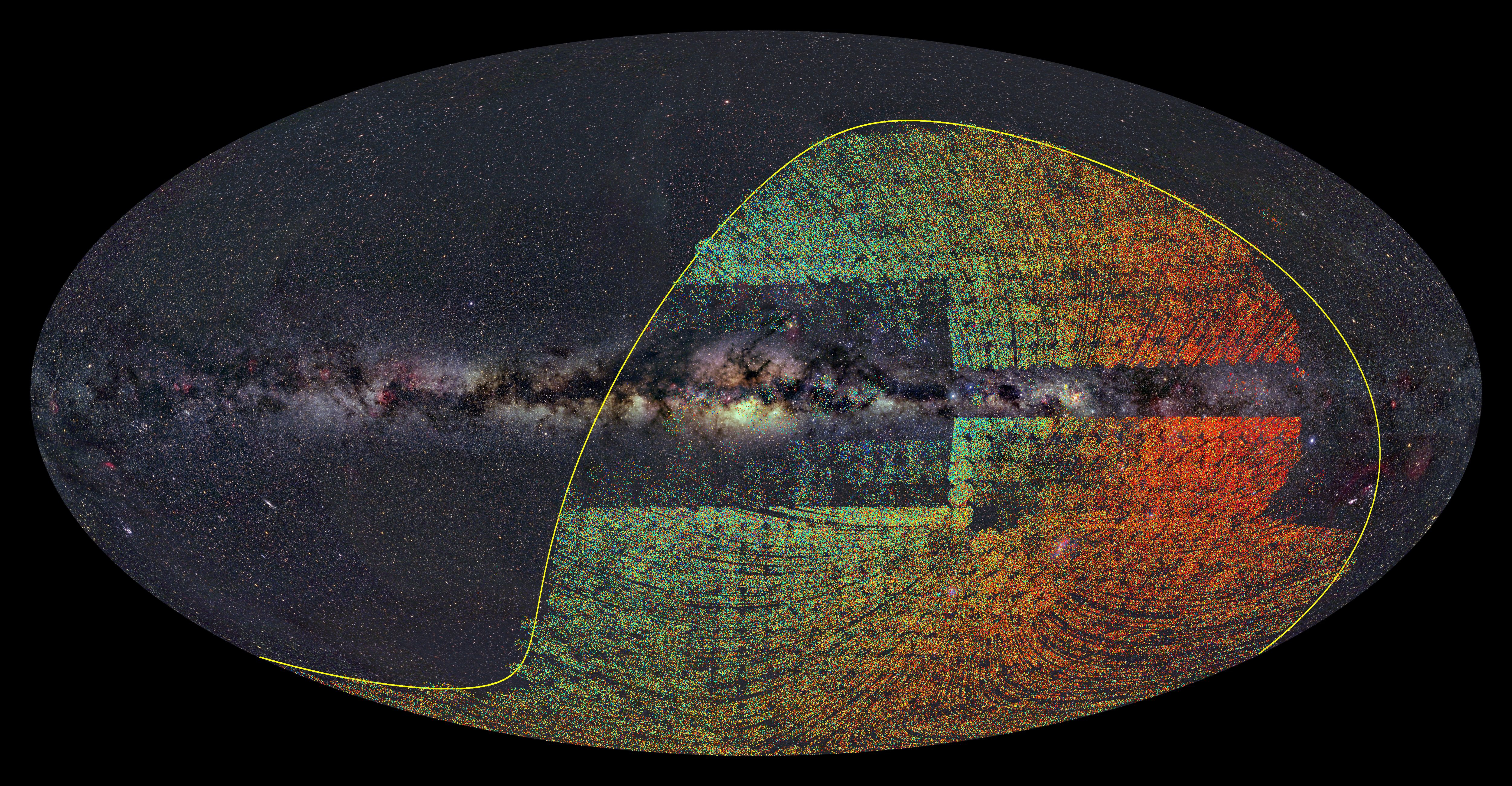}
\caption{(online colour at: www.an-journal.org) Aitoff projection of the fields covered by the RAVE survey as of June 2010. The yellow line marking the celestial equator. Color encoded is the radial velocity of each star in the field. Clearly visible is the dipole signal caused by the motion of the Sun with respect to the local standard of rest.}
\label{rave}
\end{figure}

Substructure, if at all present, is however far more difficult to identify in galactic disks. Owing to their high density and short orbital time scales, substructure in configuration space is dissolved by phase-mixing in a very few orbital time scales. In velocity space, these features are considerably longer lived. Examples for such substructure in the solar neighborhood are the moving groups first identified by Eggen (1971). While they are commonly thought to be resonant features e.g. caused by an inner bar, an accretion origin of at least some of them cannot be excluded (Navarro et al. 2004).

In this article I will discuss the structure of the thin and thick disk of the Milky Way in the light of the advent of large spectroscopic stellar surveys of Milky Way over the past few years. The selection of topics and the results presented are mostly based on recent results obtained with in the RAVE collaboration (Steinmetz et al. 2006).

\section{Spectroscopic surveys of the Galaxy}

Spectroscopic surveys are paramount to identify the debris of dissolved substructure, either by its coherency in phase space or by its chemical signatures. However, despite the importance of stellar spectroscopy, by 2005 radial velocities (RVs) for only some 50\,000 stars and even fewer stellar spectra had become available in the public databases of the Centre de Donne´es Astronomiques de Strasbourg (CDS), surprisingly few compared to the more than 1 million galaxy redshifts measured by the same epoch. The situation has changed considerably, however, in the past seven years. The public release of the Geneva Copenhagen Survey (GCS, Nordstr\"om et al. 2004) added RVs for 16\,682
nearby dwarf stars, Famaey et al. (2005) published 6691 RVs for apparently
bright giant stars. Both catalogs were part of the Hipparcos follow-up campaign.
The SEGUE campaign as part of the Sloan Digital Sky Survey delivering ${R=1800}$ spectra for 240\,000 stars (Yanny et al. 2009). This sample is complemented by further 120\,000 stars of the SEGUE-2 campaign. The SEGUE spectroscopic footprint corresponds to 212 fields. Owing to its magnitude range $14<g<20.3$, SEGUE was primarily designed to probe the Galactic halo with each field probing the Milky Way out to distances of 100 kpc.

\begin{figure}
\vskip-4mm
\includegraphics*[width=80mm]{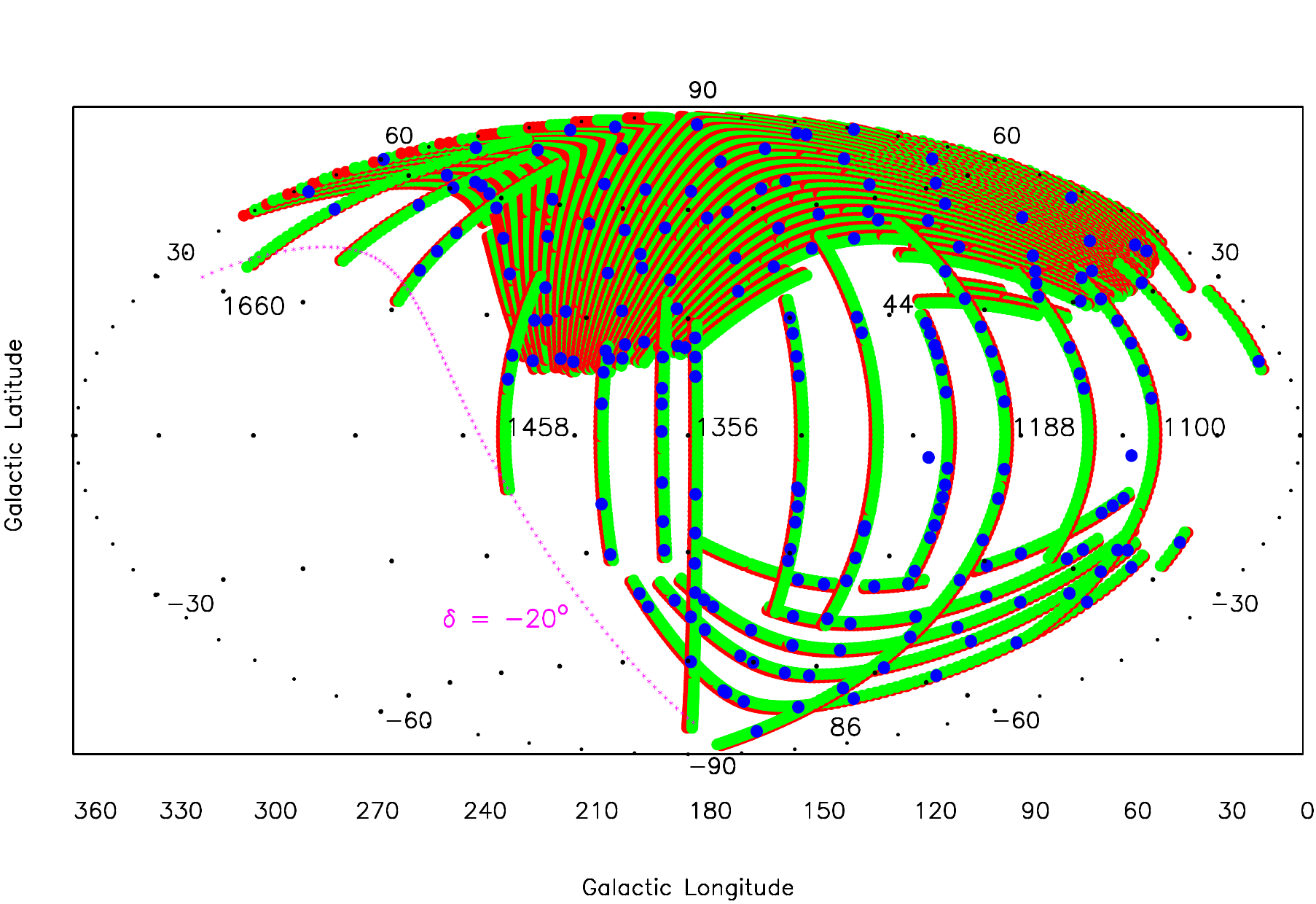}
\caption{(online colour at: www.an-journal.org) The SEGUE survey ($l, b$) in
Aitoff projection, centered on the Galactic anticenter. The line marking the
southern limit of the telescope observing site $\delta = 20\degr$ is
indicated in magenta. Red and green filled areas represent South and North SDSS
and SEGUE strips, respectively (from Yanny et al. 2009).}
\label{segue}
\end{figure}

\begin{figure}
\hskip-2mm
\includegraphics[width=64mm,angle=-90]{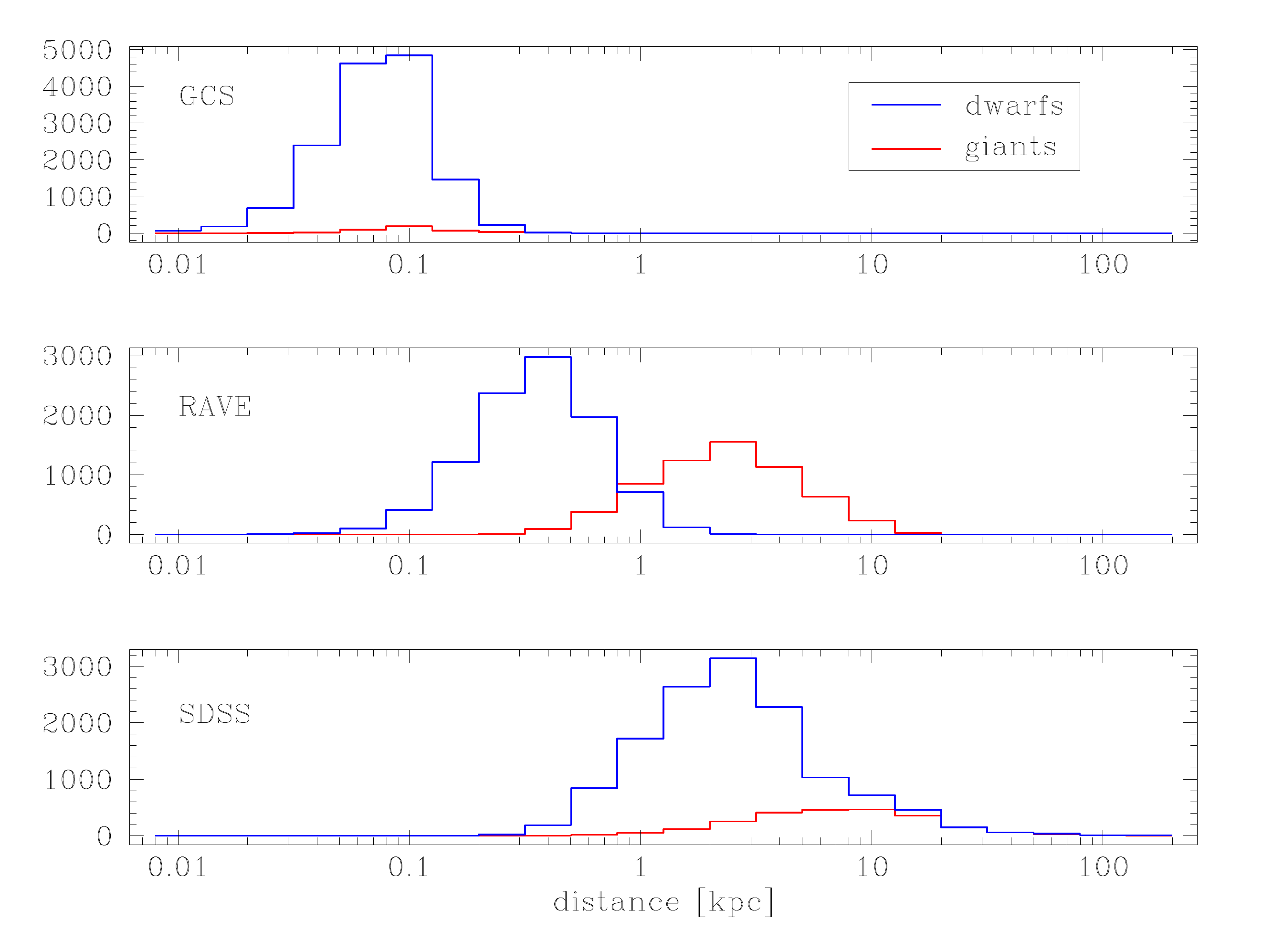}
\vskip-2mm
\caption{(online colour at: www.an-journal.org) Distance distribution of stars in the GCS (\emph{top}), 
RAVE (\emph{middle}), and SEGUE (\emph{bottom}) surveys. Dwarfs are shown in blue, giants in red.}
\label{surveys}
\end{figure}

In contrast to the deep pencil beam nature of SEGUE (Fig. \ref{segue}), the
Radial Velocity Experiment (RAVE) aims at brighter targets covering the southern
hemisphere, excluding regions covered by the galactic bulge and the denser disk
regions (Fig. \ref{rave}). RAVE uses the 6dF multi-fiber spectroscopic facility
at the UK Schmidt telescope of the Australian Astronomical Observatory in Siding
Spring, Australia and is a magnitude limited spectroscopic survey in the range
${9< I< 13}$. The wavelength range of 8410 to 8795 \AA\ at an average resolution
of $R=7500$ overlaps with the photometric Cousins $I$ band. The survey probes
both the nearby and more distant Galaxy. Typical distances for K0 dwarfs are
between 50 and 250 pc, while the O-type giants are located at distances of 0.7 to 3
kpc (Fig.~\ref{surveys}). The most distant RAVE stars, a few luminous blue variable stars, 
are actually members of the Large Magellanic Cloud.
RAVE observation commenced in April 2003 at a rate of 7 nights per lunation.
With the completion of the 6dF Galaxy Redshift Survey (Jones et al. 2004) in
August 2005, this rate increase to 20--25 scheduled nights per lunation. The
original setup consists of two field plates (complemented by a third plate in
2009) with robotically positioned fibers in the focus of the UK Schmidt
telescope. A field plate covers a 5\fdg7 field of view and feeds light to up
to 150 fibers each with an angular diameter of 6\farcs7 on the sky. In order to
avoid chance superpositions with target stars when using such wide fibers,
regions close to the Galactic plane, near the galactic bulge, or dense stellar
clusters are avoided. By April 2012 RAVE has amassed some 530\,000 spectra for
450\,000 unique stellar targets. Data products of RAVE in several data releases
include radial velocities (Steinmetz et al. 2006), stellar parameters (Zwitter
et al. 2008; Siebert et al. 2011b), distances (Breddels et al. 2010; Zwitter et
al. 2010; Burnett et al. 2011) and abundances of up to seven chemical elements 
(Mg, Al, Si, Ca, Ti, Fe, and Ni) (Boeche et al. 2011).

Figure~\ref{surveys} shows the distance distribution of targets in the GCS, RAVE, and SEGUE surveys. The composition of the RAVE survey seems to be complementary in an almost ideal fashion: The dwarf stars of RAVE extends the GCS sample to nearly 1 kpc. RAVE giants probe the Milky Way out to about 10 kpc, about the same range as the one probed by SEGUE dwarfs. SEGUE giants finally probe the galactic halo out to 100 kpc. 

\section{The Galactic thin and thick disk as seen by spectroscopic surveys}

Ever since its first postulation by Gilmore \& Reid (1983), the separation of
Milky Way disk into a thin and a thick disk has been plagued by selection
effects and biases. Recently, Bovy et al. (2011) argued, based on mono-abundance sub-populations of the Milky Way disk using SEGUE data, that there is actually no thin disk/thick disk dichotomy  and that the transition between thin and thick disk is rather a continuum of disks with an (abundance dependent) scale-length. This result is, however, difficult to bring in agreement with a similarly spirited decomposition in kinematical subcomponents towards the Galactic Pole using RAVE data, which exhibits a clear bimodality (Veltz et al. 2008). Clearly further work is needed to settle this discrepancy employing in parallel kinematical and chemical information.

The following two studies in this context are based on the RAVE survey data, namely (i) the vertical structure of the Milky Way based on purely kinematical data, and (ii) the attempt to separate the disk in distinct populations based on kinematical and abundance data, following the study of Gratton et al.~(2003). The section then concludes with a discussion of possible formation scenarios for the thick disk.

\subsection{Trends in disk kinematics with galactocentric radius and vertical height}

\begin{figure}
\includegraphics[width=40mm]{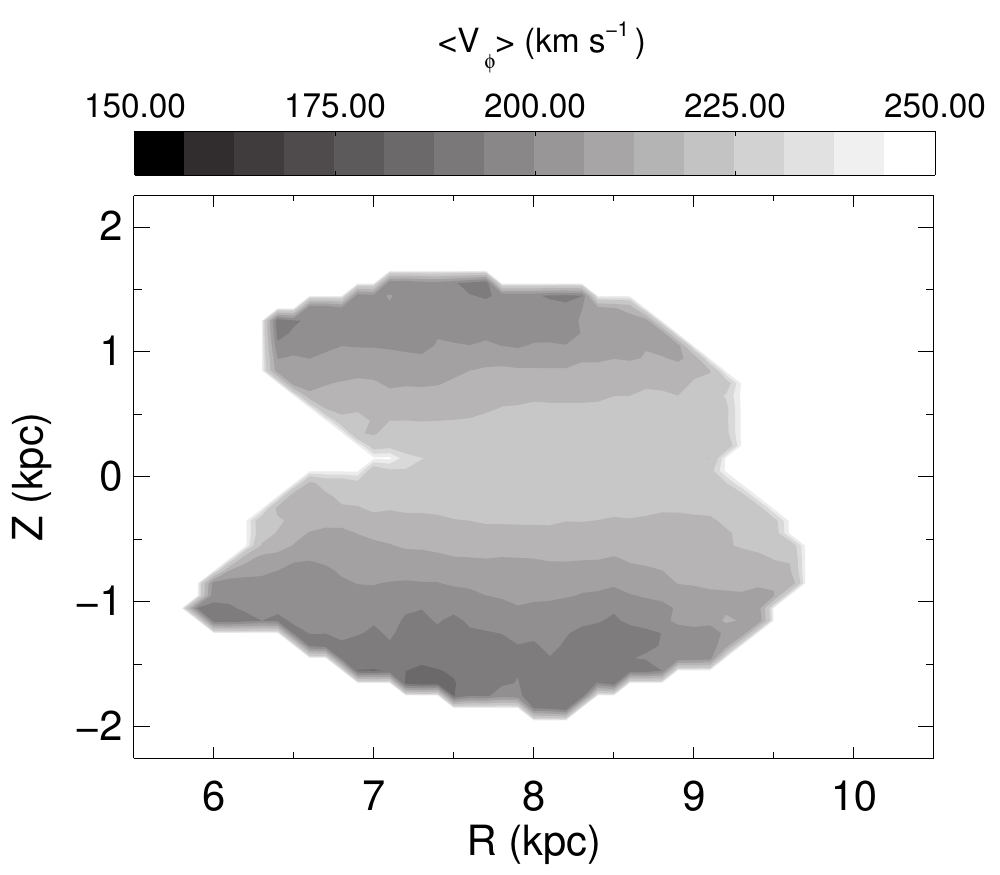}\includegraphics[width=40mm]{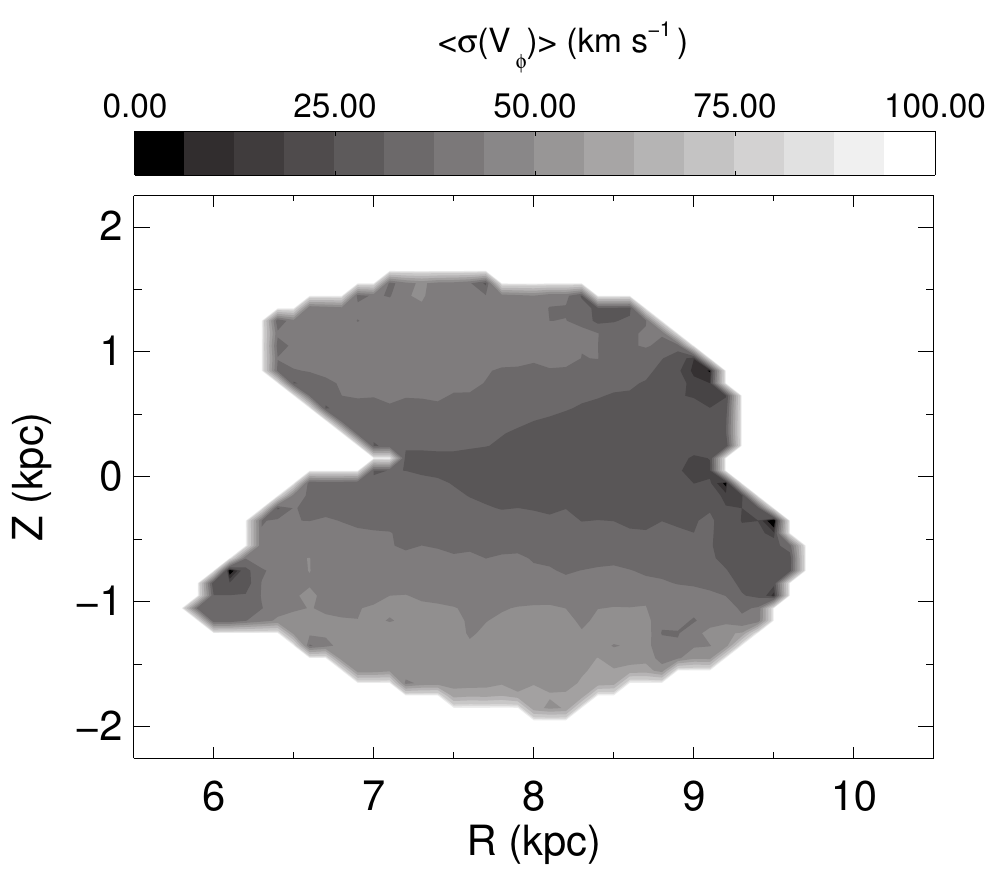}
\caption{Average orbital velocity $V_{\phi}$ (\emph{left}) and velocity dispersion $\sigma_{\phi}$ 
(\emph{right}) in the $R$-$z$ plane.}
\label{diskstruc}
\end{figure}

The vertical and radial structure of the extended solar neighborhood is illustrated in
Fig. \ref{diskstruc}. It shows the basic trends in the average orbital velocity
and the velocity dispersion of stars in the RAVE sample (Williams et al., 2012,
in preparation). From the Jeans equations, one expects a decrease in the average
orbital velocity as velocity dispersion increases (see e.g., Eq. 4-228 of
Binney \& Tremaine 2008). With increasing height above the Galactic plane $Z$ the sample becomes less dominated by thin disk stars and more by thick
disk stars. This can be seen by an increase in the velocity dispersion $\sigma_{\phi}$ (right panel), which goes along with a corresponding 
increase in the lag in $V_{\phi}$ (left panel). The results are consistent across the various distance methods
(e.g. Zwitter et al. 2010 vs. distances from red clump stars), and with the various proper
motion sources. Close to the plane, between $0.5 < Z < 0.5$
there is some hint that there is an increasing lag
in $V_{\phi}$ associated with the spiral features in the $R$-$\phi$ plane. Further
from the plane, such an association is less clear.

\subsection{Identifying Galactic components by combining chemical and kinematic data}
 A first study of the 
kinematic-abundance properties of the Milky Way thin and thick disks using RAVE 
has been recently carried out by Karatas et al.~(\cite{karatas12}). In that work the authors used 
a sample of $\sim$4000 main-sequence stars (with ${\log g >3}$) from the second 
RAVE data release and classified the Galactic disk populations according to their
distribution on the $V_{\rm rot}$-[M/H] plane, using the so-called X stellar 
population parameter defined by Schuster et al.~(\cite{schuster93}). 
Although the authors successfully reproduced the mean kinematic and metallicity 
properties of the thin- and thick-disk populations as compared to previous works 
(e.g., Veltz et al.~\cite{veltz08}), their results strongly depended on the classification 
method, which makes use of the very same kinematic and metallicity parameters. 
Based on SEGUE data, Lee et al.~(\cite{lee11}) demonstrated a strong trend of the orbital eccentricity 
with metallicity, while the eccentricity does not change with metallicity for the thin-disk subsample. 

In Boeche et al.~(2012), the RAVE data set was used to disentangle Galactic stellar populations, such as thin disk, thick 
  disk and halo based on their chemical abundances and on their orbital parameters, in particular orbital velocity, orbital eccentricity and maximum height above the galactic plane. A sample of giant stars was selected that have high signal-to-noise spectral measurements from the RAVE chemical catalogue. Following the analysis of Gratton et al.~(2003) using high-resolution spectroscopic data, the sample was subdivided into three populations, using somewhat modified selection criteria compared to Gratton et al:

\begin{enumerate}
\item Stars with ${V_{\rm rot}>40}$ km\,s$^{-1}$ and ${R_{\rm a}<15}$~kpc. They 
comprise the  {\em dissipative collapse component}.
This sample includes part of the thick disk, as well as of the halo. 

\item Non-rotating or counter-rotating stars. These stars satisfy the constraint ${V_{\rm rot}\le40}$~km\,s$^{-1}$.  This population includes part of
the halo, as well as accreted debris (which do not share the rotation of the
disk components).  This component is identified as the {\em
accretion component};

\item Stars whose orbits have low eccentricity and low inclination on the
Galactic plane.  This population satisfies the constraint $e < 0.2$ and
$Z_{\rm max}<1\,$kpc, where $Z_{\rm max}$ is the maximum 
height attained above the Galactic plane and $e$ is the orbit eccentricity. 
This is referred to as the {\em thin disk component}.

\end{enumerate}

\begin{figure}
\includegraphics[width=80mm]{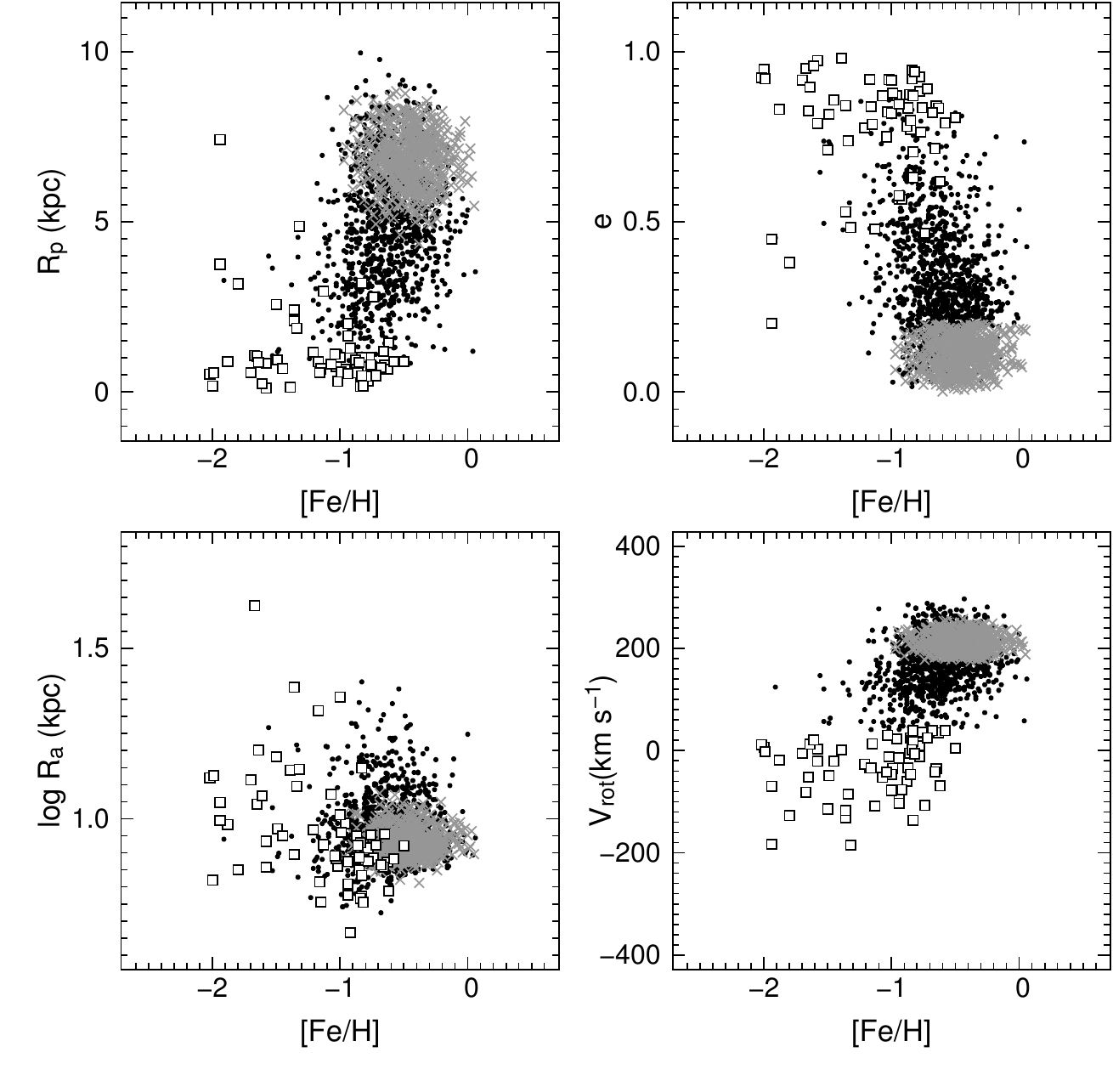}
\vskip-2mm
\caption{Perigalacticon, $R_{\rm p}$, eccentricity, $e$, apogalacticon, $R_{\rm a}$, and
rotational velocity, $V_{\phi}$ as a function of [Fe/H] for the three
subsamples. The thin disk component
(grey crosses), the dissipative component (black points), 
and the accretion component (open squares) are shown (from Boeche et al. 2012). This figure correspond to
Fig.~4 of Gratton et al. (2003).}
\label{thick}
\end{figure}

The result of such a decomposition is shown in Fig. \ref{thick} which clearly identifies these three kinematically identified populations 
also with respect to their abundance patterns. The kinematically selected  thin disk and dissipative (mostly thick disk) samples  have similar
distance and eccentricity distributions as those reported by Lee et
al.~(2011) for a G-dwarf SEGUE sample where the thin and thick
disk stars were selected based on a pure chemical criterion.
The thin disk, dissipation and accretion components
partially overlap in several parameters. Such overlaps
are to be expected if the processes currently debated in the literature
(accretion, heating and stellar migration) are at play during
the evolution of the Milky Way. This is particularly true for thin
and thick disk (the latter is identified with the dissipation component)
which kinematically overlap one another in a way that makes difficult (if not impossible) to find selection criteria capable
of disentangling them. 

\subsection{Origin of the thick disk}
\begin{figure}
\includegraphics[width=40mm]{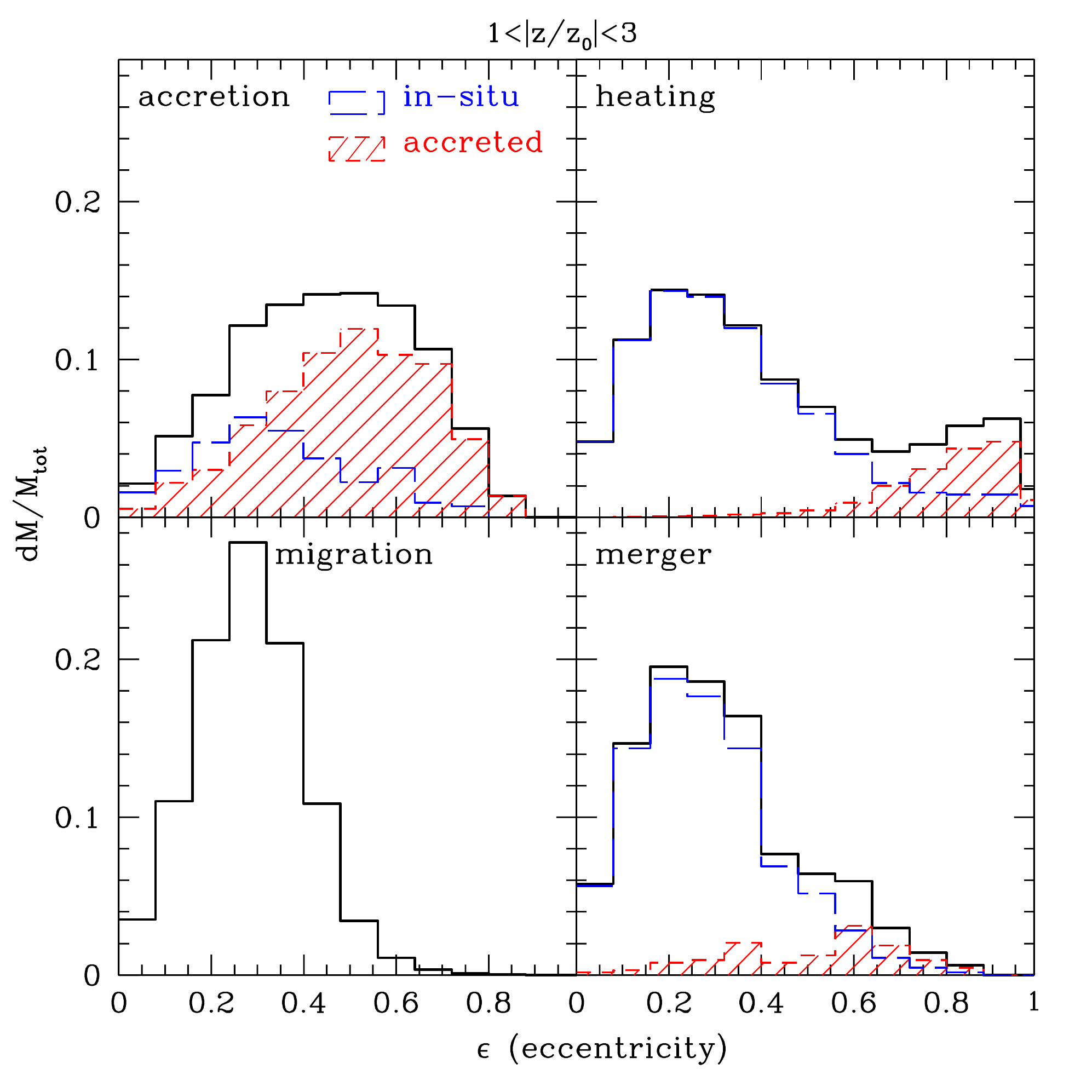}\includegraphics[width=40mm]{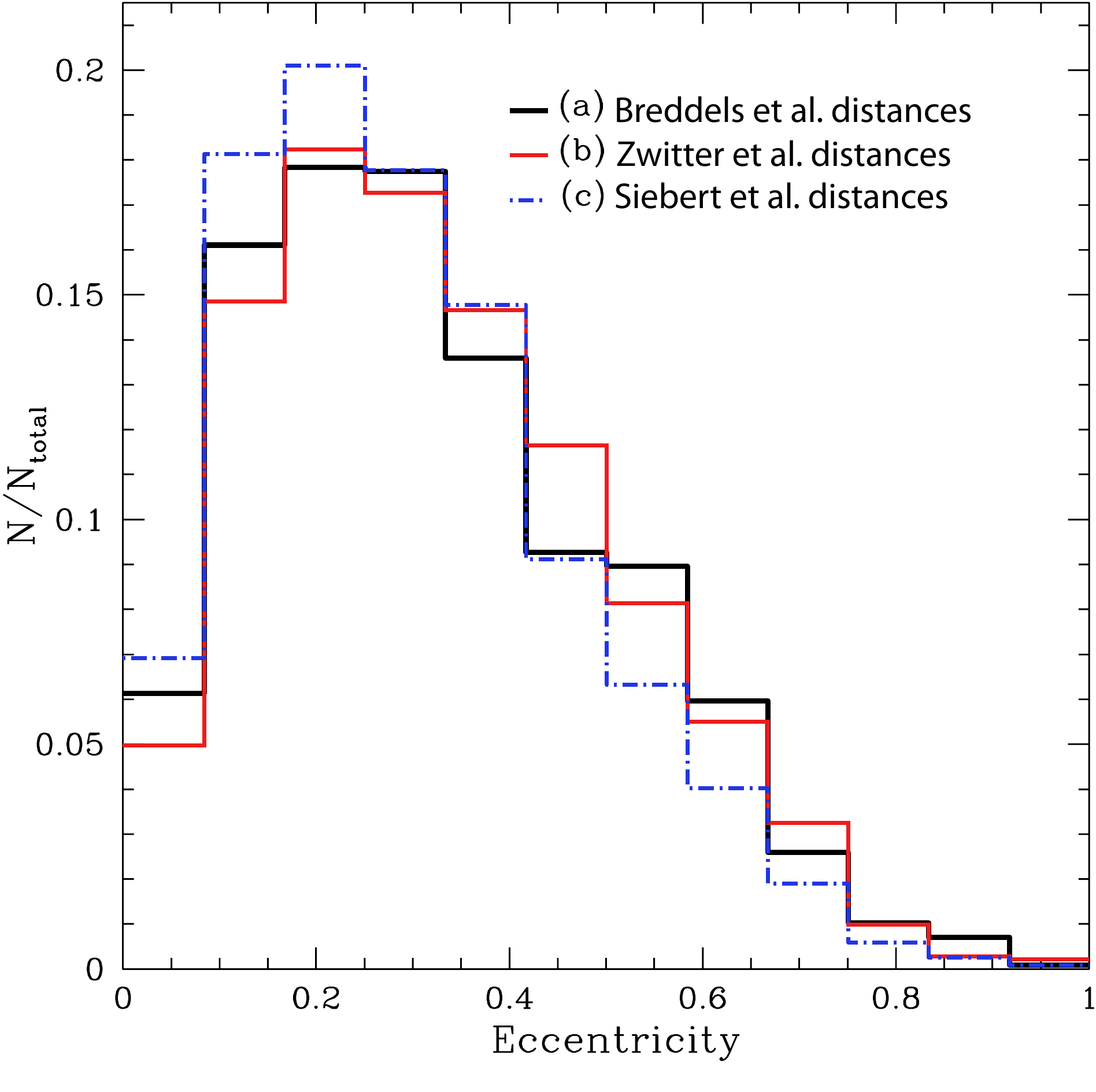}
\caption{(online colour at: www.an-journal.org) \emph{Left}: comparison of the eccentricity distributions of each thick disc
formation model for stars in the range 1--3 (thick disk) scale heights and
cylindrical distance between 2 and 3 disk scale lengths. The thick solid black line shows the
total distributions, while blue-empty and red-shaded histograms
distinguish between stars that formed in the disk (in situ) from those that were accreted (from Sales et al. 2009). 
\emph{Right}: eccentricity distributions of RAVE thick-disk samples using (a) the original distances, 
(b) the Zwitter et al. distances and (c) the clump sample (from Wilson et al. 2011).}
\label{thickform}
\end{figure}

Several scenarios that could give rise to the formation of a thick disk are currently discussed in the literature. These include (i) accretion from disrupted satellites (Abadi et al.~2003), (ii) heating of a pre-existing thin disk by a minor merger (Quinn, Hernquist \& Fullagar 1993;
Kazantzidis et al. 2008; Villalobos \& Helmi 2008), (iii) radial migration (e.g. Roˇskar et al. 2008; Sch\"onrich
\& Binney 2009), and (iv) in situ star formation triggered by star formation during gas-rich mergers (e.g. Brook et al. 2005; Bournaud, Elmegreen \&
Elmegreen 2007). Sales et al.~(2009) demonstrated that the eccentricity distribution for each of the aforementioned formation scenarios is different: a prominent peak at low eccentricity
is expected for the heating, migration and gas-rich merging scenarios, while the
eccentricity distribution is broader and shifted towards higher values for the
accretion mode (Fig. \ref{thickform}, left). 

In Wilson et al. (2011) a sample of RAVE stars with
${V_y > 50}$ km\,s$^{-1}$ (in order to exclude contamination by halo stars) and ${1 <
|z|/\mbox{kpc} < 3}$ were analyzed with respect to their eccentricity
distribution, closely mimicking the selection criteria used in Sales et
al. (2009). Furthermore, several distance measures were employed to check for
systematic effects  (Siebert et al. 2008; Breddels et al. 2010; Zwitter et
al. 2010). The resulting eccentricity distribution (Fig. \ref{thickform}, right)  exhibits a  triangular
shape with a peak at low eccentricities. Such a distribution is quite different to the one expected in case of a pure accretion scenario. It also lacks the prominent secondary
peak at high eccentricities like that evident in the heating model. The migration and merger models seem to be in better agreement with the data. In particular the gas rich merger model exhibits the characteristic triangular shape, making it the model that is  most consistent with the RAVE data. It is reassuring, that an analysis performed based on SEGUE data (Dierickx et al. 2010) and using a abundance defined subsample of RAVE stars (Ruchti et al. 2011) come to similar conclusions.

\section{Beyond simple thin disk/thick disk models}

Besides the difficulties in clearly defining various Galactic components, deviations are also expected in their overall shape and structure. Two examples are demonstrated in this section, namely deviations from axisymmetry and kinematical substructure in the disk components.

\subsection{The non-axisymmetric disk}

There are good reasons to assume that the disk of the Milky Way exhibits signs of non-axisymmetry. Such asymmetries are frequently observed in external galaxies (Bosma 1978), and there is an ample portfolio of processes that can be made responsible for such asymmetries, e.g., effects of a central bar, of spiral arms or of a triaxial dark matter halo. In the local solar neighborhood, asymmetries of the disk can be identified by determining the two asymmetric Oort's constants $C$ (radial shear) and $K$ (divergence), defined as
\begin{eqnarray}
2C = \left.\frac{\partial V_x}{\partial x}\right|_0 + \left.\frac{\partial V_y}{\partial y}\right|_0,\nonumber \\
2K =\left.\frac{\partial V_x}{\partial x}\right|_0- \left.\frac{\partial V_y}{\partial y}\right|_0 .
\end{eqnarray}

However, measurements of C and K thus far has given inconclusive results (e.g.
Kuijken \& Tremaine 1994; Olling \& Dehnen 2003). Based on RAVE data, Siebert et
al.~(2011{b}) were able to constrain the Oort constant $C$ to between --9.1 and
--13.2 km\,s$^{-1}$\,kpc$^{-1}$, the Oort constant $K$ to between 4.1 and 6 km\,s$^{-1}$\,kpc$^{-1}$,
respectively, resulting in a radial velocity gradient $\partial V/\partial
R=K+C$ in the solar neighborhood of --3 to --10 km\,s$^{-1}$\,kpc$^{-1}$. In order to
exclude that the qualitative conclusion of the existence
of non-zero velocity gradient depends on systematic trends in the estimates of
the distances, the analysis was repeated under the assumption that all distances were wrong
by a factor $f$. For $f$ between  0.8 and 1.2 the result of $C+K \ne 0$ could be confirmed.

\begin{figure}
\vskip-8mm
\includegraphics[width=83mm]{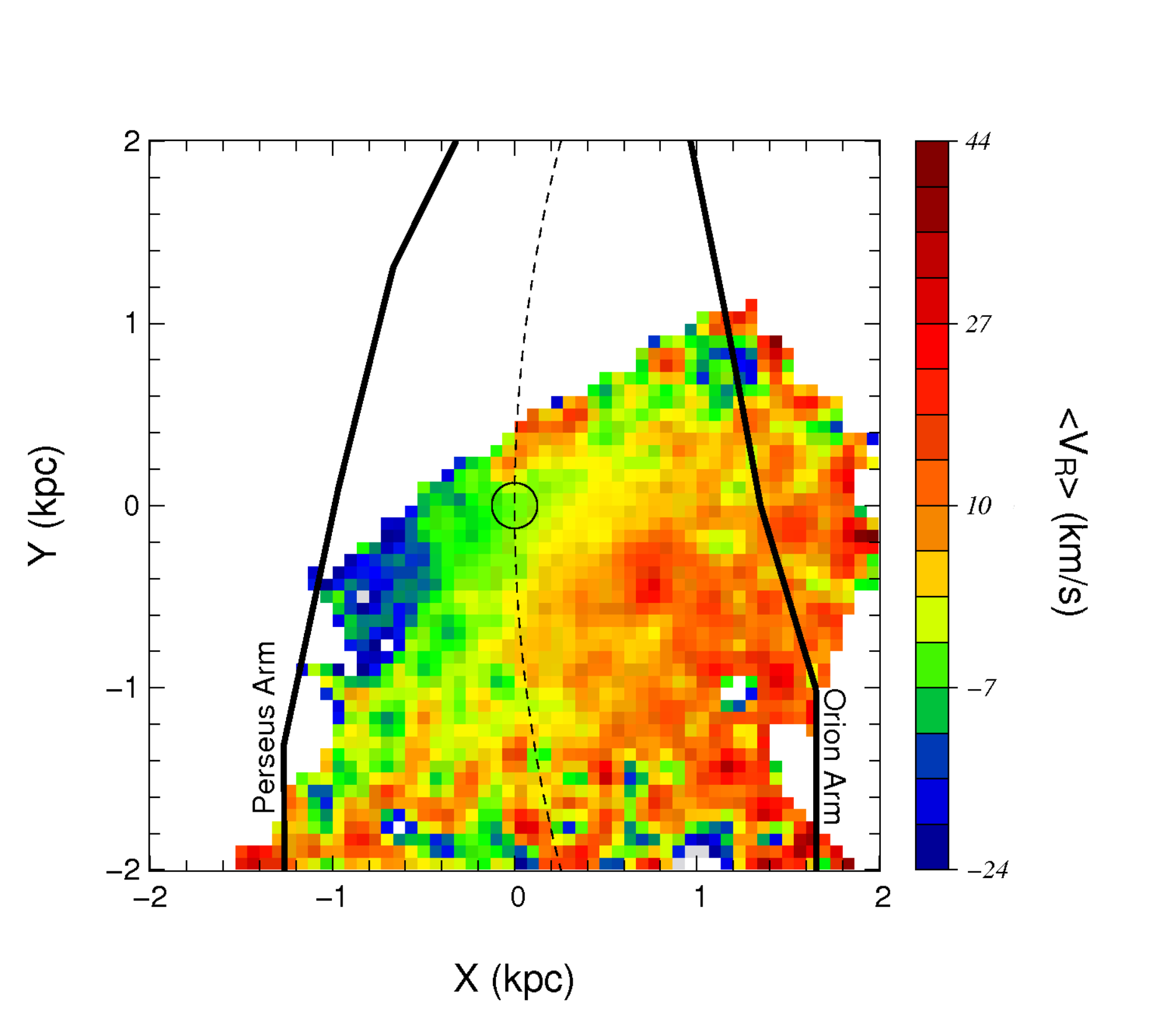}
\vskip-2mm
\caption{(online colour at: www.an-journal.org) Mean velocity field $\langle V_R\rangle$ assuming $R_0 = 8$ kpc and $v_{{\rm c},0} = 220$ km\,s$^{-1}$. The locations of the nearest spiral
arms are indicated using the CO map from Englmaier et al.~(2010). The open circle delimitates a sphere 125 pc in radius around the
Sun. The map is smoothed over 3 pixels to highlight the mean velocity trends.
The maps are 60${\times}$60 bins in size between --2 and 2 kpc
along each axis. $X$ increases in the direction of the Galactic centre, $Y$ is
positive towards the Galactic rotation (from Siebert et al. 2011a). \vspace{-2mm}}
\label{gradient}
\end{figure}

The velocity gradient becomes clearly visible in a 2D velocity map of the mean
radial velocity $\langle V_R\rangle$ centered around the solar circle (Fig. \ref{gradient}). 
The solar circle ($R = R_0$) is drawn as
a dashed line and the open circle delimits a zone 125 pc
in radius similar to the Hipparcos sphere. For orientation,
the location of the Sagittarius-Carina and Perseus arms are indicated
based on the CO map of Englmaier et al. (2010). The velocity of the Sun with respect to the local standard of rest is fixed to the value derived 
by Sch\"onrich et al.~(2010).
 $X$ and $Y$ increase towards the GC and in
the direction of Galactic rotation, respectively. While there is no gradient visible in the solar neighborhood (indicated by the
open circle), 
compatible with previous measurements, a clear velocity gradient
becomes noticeable outside the Hipparcos sphere.  Isocontours in $\langle V_R\rangle$  are
oriented about 50\degr\ from the direction to the GC and do
not follow a particular symmetry axis of the 
sample. The gradient shows some dependency on the assumed value of $R_0$ and $v_{{\rm c},0}$, however trying to
minimize or even eliminate the gradient while keeping the ratio $v_{{\rm c},0}/R_0$ constant, would require rather unrealistic values for the Galactic parameters with ${R_0 < 4}$ kpc and ${v_{{\rm c},0} < 126}$ km\,s$^{-1}$.

In a follow-up study, Siebert et al.~(2012) applied the traditional density
wave theory and analytically modeled the radial component of the two-dimensional
velocity field. Provided that the radial velocity gradient is caused by relatively long lived
spiral arms that can affect stars substantially above the plane, this analytic
model provides new independent estimates for the parameters of the Milky Way spiral
structure. The analysis favors a two-armed perturbation with the Sun close to the
inner ultra-harmonic 4:1 resonance, with a pattern speed $\Omega_{\rm p}\approx 18.6$ km\,s$^{-1}$\,kpc$^{-1}$ and 
a small amplitude $A \approx 0.55$ of the background potential.

\subsection{Identifying substructure in velocity space: the Aquarius stream}

One of the major objectives of spectroscopic Galactic surveys is the identification of chemically and kinematically coherent structures. The presence of large vertical streams in the solar neighborhood can be excluded based on symmetry considerations (Seabroke et al.~2008): A stream penetrating  the Galactic disk  from the NGP  to the SGP would
exhibit vertical velocity -W both above and below
the plane while +W both above and below the plane falling from the
SGP to the NGP. A symmetry test between the two sides of the W
distribution, if sensitive enough to the number of stream stars, will
find the asymmetry caused by a single stream. The absence of such a clear asymmetry limits the number of stars in vertical streams to be  smaller than the Virgo overdensity. 
Regarding smaller structures, Antoja et al.~(2012) recently used the RAVE data to demonstrate, that some of the Eggen moving groups are non-local features and can be identified at least up to 1 kpc from the Sun in the direction of anti-rotation, and also at 700 pc below the Galactic plane.

\begin{figure}
\vskip-2mm
\includegraphics[width=80mm]{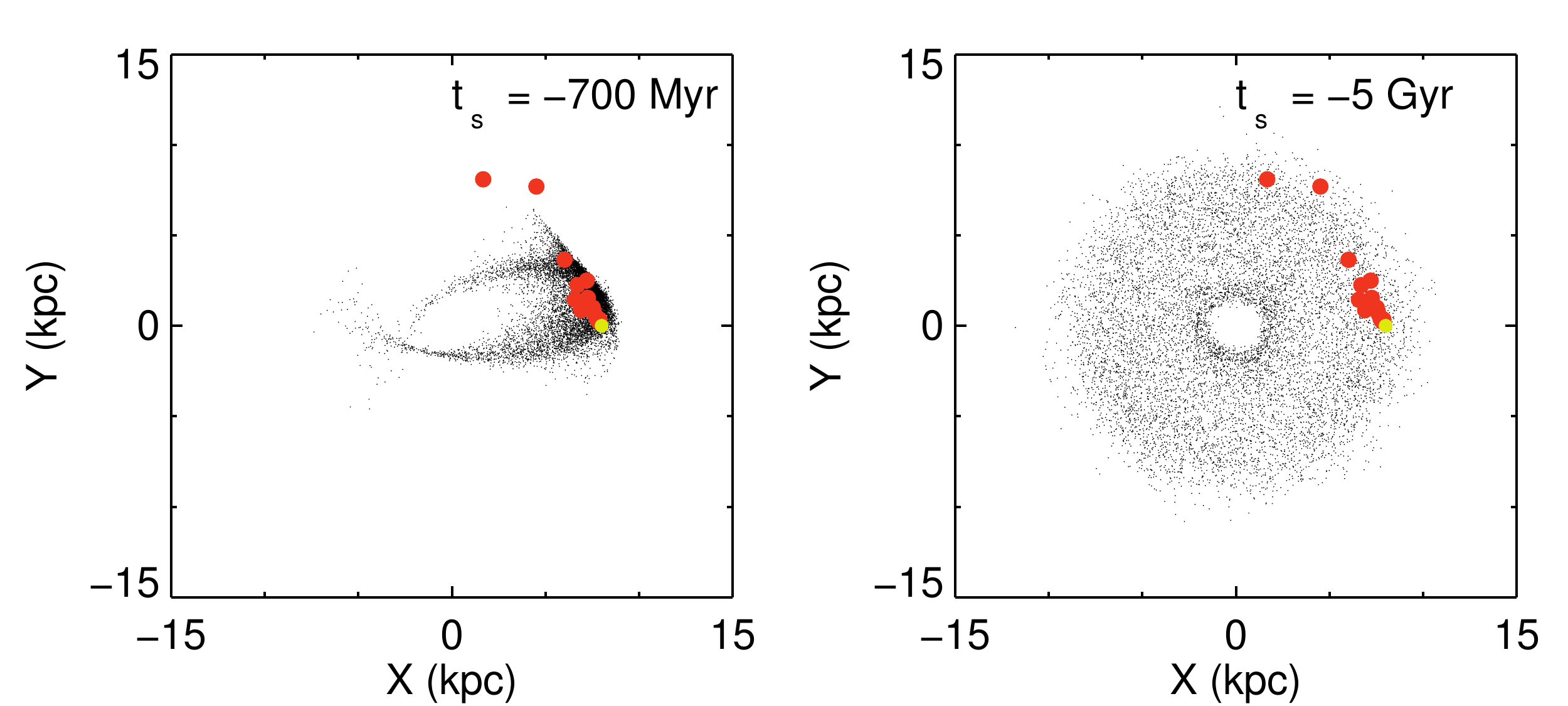}
\includegraphics[width=80mm]{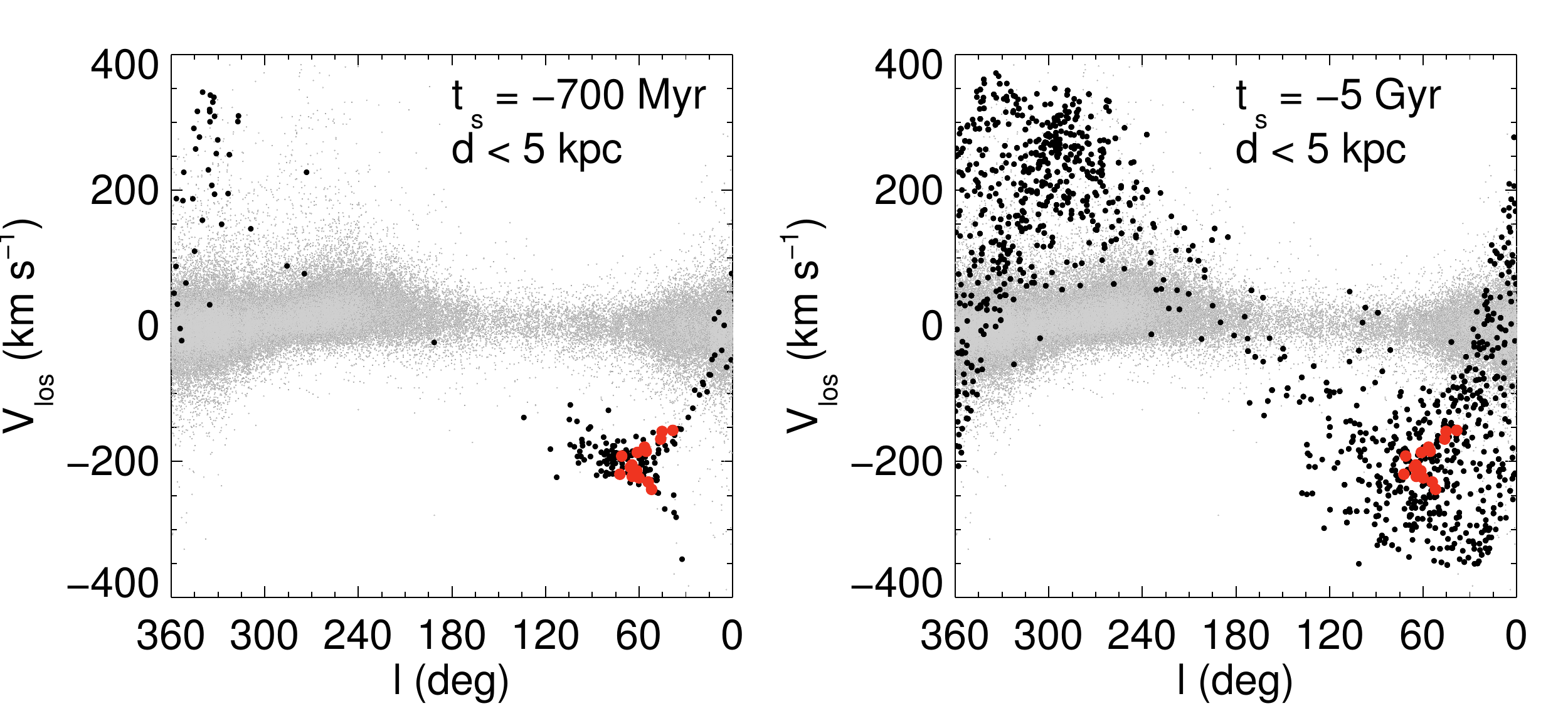}
\caption{(online colour at: www.an-journal.org) \emph{Top}: distribution of the
Aquarius members (red) and the stars of a model simulation (black) in the
$l$-$V_{\rm los}$ plane for accretion event 700 Myr ago (\emph{left}) and 5 Gyr ago (\emph{right}).
 \emph{Bottom}: distribution of the tidal debris in the orbital plane for the two accretion times, respectively. The background RAVE distribution is given in grey (from Williams et al. 2011).  \vspace{-1mm}}
\label{aquarius}
\end{figure}

A genuinely new structure in the Milky Way disk is the so-called Aquarius stream. It was identified using the RAVE catalogue and thus far consists of 
15 members of the stream lying between ${30\degr < l < 75\degr}$ and ${-70\degr <
b < -50\degr}$, i.e., in the constellation of Aquarius, with heliocentric
line-of-sight velocities ${V_{\rm los} \approx -200}$ km\,s$^{-1}$ (see Fig. \ref{aquarius}). The Aquarius members are outliers in the radial velocity distribution, and the overdensity is statistically significant when compared to mock samples created with both the Besancon and Galaxia Galaxy models. The
metallicity distribution function and isochrones suggest the stream consists of
a 10 Gyr old population with [M/H] ${\approx -1.0}$. Using a simple dynamical model of a dissolving satellite galaxy the localization of the stream is accounted for. It is dynamically young and therefore likely
the debris of a recently disrupted dwarf galaxy or globular cluster. The Aquarius stream is thus a
specimen of ongoing hierarchical Galaxy formation, rare for being right in the solar suburb.

A follow-up campaign performed to get abundances of six member stars of the
Aquarius stream confirmed the estimated age (12 Gyr) and metallicity
([Fe/H] $=-1$) for these stars (Wylie-deBoer et al. 2012). In order to investigate whether the stream represents the debris
of a disrupted dwarf galaxy or a disrupted globular cluster, the 
[Ni/Fe]-[Na/Fe] plane provides a good diagnostic as globular cluster stars and dwarf
spheroidal galaxy stars are well separated in this plane. Based on this diagnostic, the Aquarius stream
stars lie unambiguously in the globular cluster region. Furthermore, a tentative northern counterpart of 
Aquarius stream could be identified in SDSS.

\section{Future opportunities}

Spectroscopic surveys such as GCS, RAVE and SEGUE are just the first in a number
of large surveys designed to unravel the structure of the Galaxy. Scheduled for
launch in 2013, the ESA cornerstone mission Gaia will survey the position,
proper motion and distances of up to a billion stars. The Radial Velocity
Spectrometer (RVS) on board of the Gaia satellite will take spectra for up to
100 million stars. The spectral range is the Ca-triplet region, similar to the
RAVE survey, the resolution for the brightest 1 million stars ($V< 11$)
corresponds to $R=11\,000$, for fainter targets the resolution is binned down by a factor of three.

The Gaia mission is complemented by a number of ground based surveys. The Gaia-ESO survey is a 300-night campaign at ESO's Very Large Telescope. It is a public spectroscopic survey, targeting $10^5$ stars, systematically covering all major components of
the Milky Way, from halo to star forming regions. For a subset of $10^4$ field stars with ${V<15.5}$ detailed abundances will be obtained by UVES spectroscopy for at least 12 elements (Na, Mg, Si, Ca, Ti, V, Cr, Mn, Fe, Co, Sr, Zr, Ba) and for several other elements (including Li) for
more metal-rich cluster stars.

HERMES, to begin observations in 2013, is a spectroscopic campaign using the
3.9m telescope of the Australian Astronomical Observatory. In about 1000 nights
of observing time, HERMES aims to obtain precision multi-element abundances for
a million stars with $V<14$, from high S/N, $R=28\,000$ spectra in 4 wavelength windows. 

The probably most ambitious spectroscopic surveys for the foreseeable future are
BIGBOSS and 4MOST. BIGBOSS is a project planned by the US community using the 4m
Mayall telescope at Kitt Peak to run a highly multiplex (${N=5000}$) ${R=5000}$
spectroscopic survey. While the main science aim is an all-sky galaxy redshift
survey in order to constrain the properties of dark energy, BIGBOSS will also
have a  galactic archeology focus as part of its community science programme.
4MOST is currently a conceptual design study for ESO with the aim of designing a
wide-field, high multiplex fiber-fed spectroscopic survey facility on one of
ESO's 4\,m-class telescopes. One of the main science drivers for 4MOST is to
further enhance our understanding of the Milky Way via follow-up spectroscopy of
Gaia targets. 4MOST will have a large enough field-of-view ($>$3 sq.\,deg.,
goal ${>}$5 sq.\,deg.) to survey a large fraction of the southern sky in a
few years. The facility will have a high multiplex ($>$1500 fibers, goal 3000)
and a high enough spectral resolution to detect of chemical and kinematic
substructure in the stellar halo, bulge and thin and thick discs of the Milky
Way, and enough continuous wavelength coverage (at least 390--950 nm) to secure velocities of extra-galactic objects over a large range in redshift.  4MOST will run permanently on the selected telescope to perform a 5 year public survey yielding more than 10 million (goal $>$25 million) spectra at resolution ${R>5000}$ and more than 1 million spectra at ${R>20\,000}$.

\section{Summary and conclusions}

The few examples described in this review are only a very first application and illustrate the huge potential that
spectroscopic surveys of the Milky Way bear in disentangling the structure, evolution and formation history of our Milky Way. The examples already demonstrate that simple descriptions of the Milky Way disks base on two symmetric components are insufficient to account for the rich structure present in disk galaxies. Though the Milky Way  disks provide clear evidence of a tight link between the chemical and kinematical properties of its constituting stellar populations, substructure, asymmetries and tidal debris seem to considerably modify the overall picture drawn in these very first studies. Upcoming missions, in particular the Gaia astrometric mission and deep spectroscopic surveys such as 4MOST will enable us to establish a far more comprehensive view of the Milky Way and its formation history.

\acknowledgements
The work presented in this review relies heavily on results obtained with the RAVE survey, 
in particular of RAVE team members Mary Williams, Corrado Boeche and Arnaud Siebert. Funding
for RAVE has been provided by: the Australian Astronomical Observatory; the
Leibniz-Institut f\"ur Astrophysik Potsdam (AIP); the Australian National
University; the Australian Research Council; the French National Research
Agency; the German Research Foundation (SPP 1177 and SFB 881); the European
Research Council (ERC-StG 240271 Galactica); the Istituto Nazionale di
Astrofisica at Padova; The Johns Hopkins University; the National Science
Foundation of the USA (AST-0908326); the W.M.~Keck foundation; the
Macquarie University; the Netherlands Research School for Astronomy; the
Natural Sciences and Engineering Research Council of Canada; the Slovenian
Research Agency; the Swiss Naleetional Science Foundation; the Science \&
Technology Facilities Council of the UK; Opticon; Strasbourg Observatory;
and the Universities of Groningen, Heidelberg and Sydney.  The RAVE web site
is at http://www.rave-survey.org.

\end{document}